\documentclass[prb,twocolumn,showpacs,amsmath,amssymb]{revtex4}

\usepackage{amsmath}
\usepackage{amssymb}
\usepackage{graphicx}
\usepackage{color}
\usepackage{pifont}
\usepackage[pdftex,dvips]{epsfig}
\newcommand{\be}{\begin{equation}} 
\newcommand{\ee}{\end{equation}}
\newcommand{\bma}{\begin{displaymath}}
\newcommand{\ema}{\end{displaymath}}

\usepackage[sort&compress]{natbib}

\begin{document} 

\title{Structural, chemical and dynamical trends in graphene grain boundaries}

\author{Sami Malola$^1$, Hannu H\"akkinen$^{1,2}$, and Pekka Koskinen$^1$\footnote{Corresponding author}}

\email[email:]{pekka.koskinen@iki.fi}

\affiliation{NanoScience Center, Department of Physics$^1$ and Chemistry$^2$, University of Jyv\"askyl\"a, 40014 Jyv\"askyl\"a, Finland}

\date{\today}

\begin{abstract} 
Grain boundaries are topological defects that often have a disordered character. Disorder implies that understanding general trends is more important than accurate investigations of individual grain boundaries. Here we present trends in the grain boundaries of graphene. We use density-functional tight-binding method to calculate trends in energy, atomic structure (polygon composition), chemical reactivity (dangling bond density), corrugation heights (inflection angles), and dynamical properties (vibrations), as a function of lattice orientation mismatch. The observed trends and their mutual interrelations are plausibly explained by structure, and supported by past experiments.
\end{abstract}

\pacs{61.72.Mm,61.48.Gh,73.22.Pr,71.15.Nc}
\maketitle

\section{Introduction}
Real graphene has always defects, edges, point defects, chemical impurities---and grain boundaries. Grain boundaries (GBs) are topological defects, trails of disorder that separate two pieces of pristine hexagonal graphene sheets. They reside practically in any graphitic material, graphite\cite{Cervenka_JPCS_07,Cervenka_naturephys_09,cervenka_PRB_09}, soot\cite{Iijima_JPC_96,Boehman_EF_05,Muller_CatToday_05}, single- and multi-layer graphene\cite{varchon_PRB_08,Chae_AdvMat_09}, fullerenes\cite{Terrones_SC_02,Lau_PRB_07}, carbon nanotubes\cite{Charlier_AccChemRes_02,Ren_APL_99}, with varying abundance. In graphene fabrication, for instance, chemically synthesized samples tend to have more GBs than mechanically cleaved samples\cite{Geim_Science_09}. For electron mobility high purity is important and GBs are best avoided, while other properties may tolerate defects better. We refer here to strictly two-dimensional GBs (albeit not necessarily planar) which should not be confused with genuinely three-dimensional GBs in graphite.

But not are defects, impurities and GBs always something bad; they are interesting also in their own right---even useful\cite{Stankovich_Nature_06,Geim_Science_09}. After all, being extended, GBs modify graphene more than point defects. They affect graphene's magnetic, electronic, structural, and mechanical properties\cite{Neto_RevModPhys_09,Geim_Science_09}.

Considering the relevance and prevalence of graphene GBs, they ought to deserve more attention. Most related studies are on GBs in graphite and moir\'{e} patterns therein\cite{biedermann_PRB_09,gan_SS_03,simonis_SS_02,cervenka_PRB_09}, or on intramolecular junctions in carbon nanotubes\cite{yao_nature_99,ouyang_science_01,chico_PRL_96}; GBs in single- or few-layer graphene, whether experimental\cite{gu_APL_07,varchon_PRB_08,Cervenka_naturephys_09} or theoretical\cite{araujo_PRB_10}, are still scarce. Often graphitic grain boundaries are idealized by  pentagon-heptagon pairs\cite{simonis_SS_02,araujo_PRB_10}, where the pair distances determine lattice mismatch. While this ideal model works in certain occasions\cite{simonis_SS_02,araujo_PRB_10}, it does not work for GBs that are rough, corrugated, have ridges\cite{biedermann_PRB_09}, and---most importantly---are decorated by dangling bonds or reactive sites alike\cite{Cervenka_naturephys_09}.

In this paper we investigate graphene GBs beyond simple models. We explore computationally the trends in structural properties, chemical reactivity and vibrational properties for an ensemble of GBs that are \emph{not ideal}, but have certain roughness. What this roughness really means will be clarified in Sec.III.

\def\egr{\epsilon_\text{gr}}
\section{Computational method}
\label{sec:DFTB}
The electronic structure method we use to simulate the GBs is density-functional tight-binding (DFTB)\cite{porezag_PRB_95,elstner_PRB_98,koskinen_CMS_09}, and the \texttt{hotbit} code\cite{hotbit_wiki}. DFTB models well the covalent bonding in carbon\cite{malola_PRB_08,malola_PRB_08b,malola_EPJD_09}, and suits fine for our simulations that concentrates on trends. We leave the details of the method to Ref.~\onlinecite{koskinen_CMS_09} and only mention---this will be used later---that the total energy in DFTB, including the band-structure energy, the Coulomb energy, and the short-range repulsion, can be expressed as a sum of atomic binding contributions,
\begin{equation}
E_\text{DFTB} = \sum_{i=1}^N \epsilon_i.
\label{eq:etot}
\end{equation}
Here $N$ is the atom count and $\epsilon_i$ atom $i$'s contribution to cohesion energy; in pristine graphene $\epsilon_i\equiv \egr=-9.6$~eV for all atoms (DFTB has some overbinding). The quantity $\epsilon_i$ allows us to calculate energies in a local fashion: for a group of carbon atoms in a given zone $\mathcal{S}$ (set of atoms), the sum 
\begin{equation}
\Delta E=\sum_{i \in \mathcal{S}} (\epsilon_i-\egr)
\label{eq:locale}
\end{equation}
measures how much the energy of zone $\mathcal{S}$ is larger than the energy of equal-sized piece of graphene. In addition, since $\epsilon_\text{gr}=-9.6$~eV comes from $3$ equivalent bonds, the value $\epsilon_i=2/3\cdot \egr\approx -6.4$~eV infers a dangling bond for atom $i$. Hence, albeit less accurate than density-functional theory, DFTB enables straightforward analysis of local quantities, in addition to fast calculations and extensive sample statistics.

\section{Sample construction}
\begin{figure}[t!]
\includegraphics[width=1.0\columnwidth]{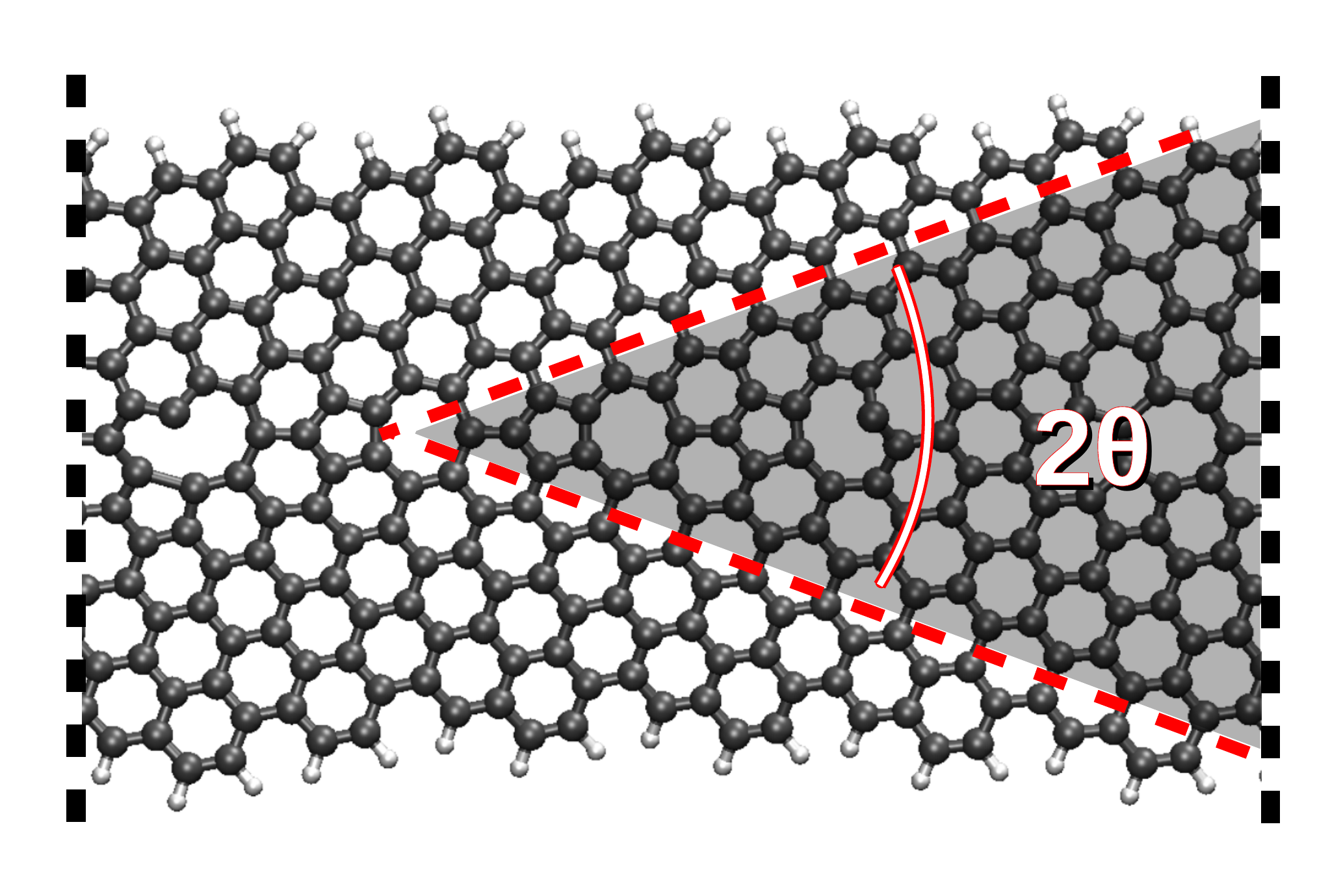}
\caption{(color online) Grain boundary sample, one out of $48$, made by merging two graphene edges with the same chirality, but different orientation and random translation along the boundary; our samples are hence characterized by chirality. In this figure chiral indices are $(11,6)$, chiral angle $\theta=20.4^\circ$. Dashed vertical lines stand for periodicity across the horizontal direction.}
\label{fig1}
\end{figure}

To investigate trends, we constructed an ensemble of $48$ graphene GBs using a procedure we now describe. (i) Cut two ribbons out of perfect graphene in a given chirality direction, with random offset. Chirality index $(n,m)$ means we cut graphene along the vector
$\mathbf{C}=n\mathbf{a}_1+m\mathbf{a}_2$, where $\mathbf{a}_i$ are the primitive lattice vectors. The cut direction can be expressed, equivalently, vy the chiral angle $\theta=\tan^{-1}[\sqrt{3}m/(m+2n)]$. (ii) Passivate ribbons' other edges with hydrogen, leaving carbon atoms on the other edges bare. (iii) Thermalize the ribbons, still separated, to $1500$~K using Langevin molecular dynamics (MD). (iv) Perform reflection operation on the other ribbon with respect to a plane along ribbon axis and perpendicular to ribbon plane. (v) Perform random translations along the ribbon direction. (vi) Merge the two ribbons' bare carbon edges with MD at $1500$~K. (vii) Cool the system to room temperature within $\sim 1$~ps. (vii) Optimize the geometry using the FIRE method\cite{bitzek_PRL_06}, and finally obtain structures like the one in Fig.\ref{fig1}. The total process takes about $3$~ps.

Our ensemble of $48$ GB samples are chosen to get representatives for all chiral angles from $\theta=0^\circ$ to $\theta=30^\circ$. Our limit for the width in cutting the ribbons is $10$~\AA. Ribbons have to be wide enough ($>9$~\AA) for faithful description of a semi-infinite graphene, and, by limiting the total number of atoms below $170$, also the periodicity of GB becomes limited below $40$~\AA. (Edges with $\theta\sim 0^\circ$ and $\theta\sim 30^\circ$, that would have short periodicity, are scarce.) Our procedure to make GBs is neither the only nor the best one, but is suitable for trend-hunting.

Sure enough, it's a downside that the procedure excludes GBs from ribbons with different edge chirality, such as merged zigzag and armchair edges. Different chirality, unfortunately, would mean different periodicity and practical problems. To have a single number, $\theta$, to identify GBs is vital for finding the trends we concentrate on. For GBs identified by two numbers, $\theta_1$ and $\theta_2$, trend-hunting would be harder; those GBs can be investigated afterwards, using the insights we learn here. Further, while the temperature $1500$~K, motivated partly by experiments\cite{biedermann_PRB_09}, allows rearrangements in merging, the heat of fusion renders initial temperatures irrelevant. We use canonical MD in the merging to allow \emph{GB formation process to be steered by intrinsic driving forces and energetics}. 

The time scale of the process is limited, as usual, in part by computational constraints. However, prolonging the process would cause no fundamental structural changes as the merging process itself is nearly instantaneous; there is no diffusion after merging and saturation of the dangling bonds, there is only annealing of the worst local defects and stress-release that causes buckling (Sec.\ref{sec:buckling}). The usual time scales in graphene growth are hence not relevant to our GB formation process. Only intrinsic roughness remains in the GB samples---roughness related to chiral angle and randomness in cutting offsets and translations.

Usually all atoms in ribbons were also bound to GBs, but in few structures (near armchair chirality) the merging process squeezed dimers out, and either inflected them out of plane or detached them completely. Since ribbons found these own ways to optimize geometries---ways not anticipated beforehand---it suggests that the design of our construction process is not dominant.

\section{Prelude: energies for free graphene edges}

\begin{figure}[t!]
\includegraphics[width=1.0\columnwidth]{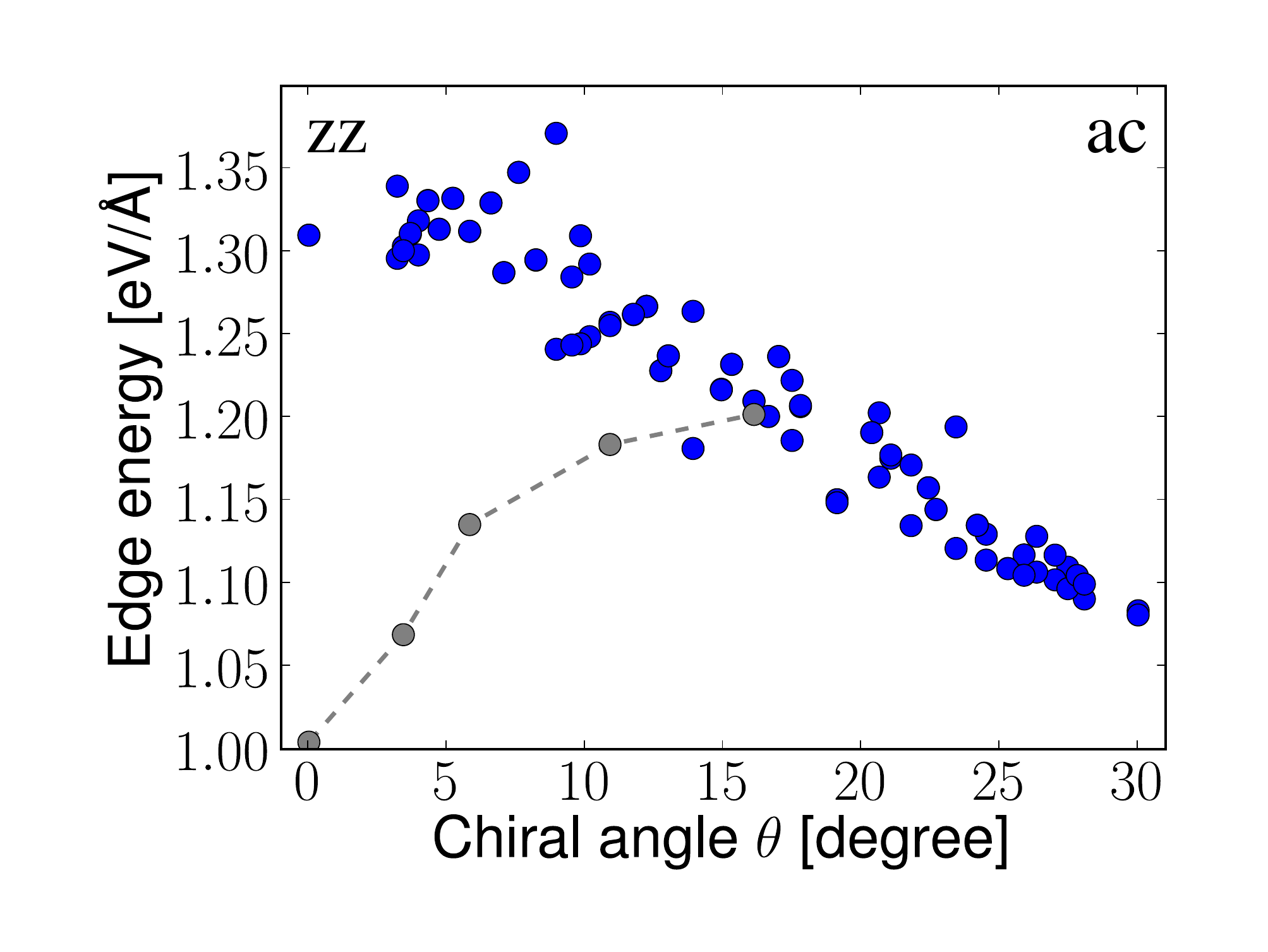}
\caption{(color online) Energies of the free edges of graphene as a function of chiral angle; zigzag edge has $\theta=0^{\circ}$ and armchair $\theta=30^{\circ}$. Free zigzag edge segments are metastable and prefer local reconstruction (two adjacent hexagons reconstruct into a pentagon and a heptagon); these so-called reczag edges are shown with gray connected symbols. The reczag segments are, however, important only for free edges and irrelevant for grain boundaries studied here.}
\label{fig2}
\end{figure}

Before going into GBs themselves, let us look at the free bare edges of graphene. Fig.\ref{fig2} shows the edge energies as a function of the chiral angle, $\theta=0^{\circ}$ meaning zigzag and $\theta=30^{\circ}$ armchair edges. To ignore the effect of hydrogen passivated edge, carbons bound to hydrogen are neglected in edge energy calculation,
\begin{equation}
\label{edge}
\varepsilon_\text{edge}=\frac{1}{L}\sum_{i \in \text{edge}} (\epsilon_i-\egr).
\end{equation}

The edge energy between zigzag and armchair varies linearly; fluctuations in energy are due to random offset in the cut, occasionally producing pentagons. The edge energy in zigzag is high due to strong and unhappy dangling bonds; in armchair the dangling bonds are weakened by the formation of triple bonds in the armrest parts\cite{koskinen_PRL_08,malola_EPJD_09}.

It was recently predicted theoretically\cite{koskinen_PRL_08} and later confirmed experimentally\cite{koskinen_PRB_09} that the zigzag edge is actually metastable, and prefers reconstruction into pentagons and heptagons at the edge, forming a so-called reczag edge. As it turned out, reczag is energetically even better than armchair. Hence, if we reconstruct the zigzag segments in edges with small $\theta$, we usually lower the edge energy, as seen in Fig.\ref{fig2} where reconstructions are added by hand. Reczag edges are, however, irrelevant for our GBs where edges are \emph{not} free, and are ignored because we want the dangling bonds to spontaneously find contact from the other merging edge. The edge energies were investigated and presented here for comparison with DFT calculations\cite{koskinen_PRL_08}. The accuracy in edge energies is better than $10$~\%, and we expect same accuracy in GB energetics.

\section{Trends in energy and structure}
\label{sec:structure}
\begin{figure}
\includegraphics[width=1.0\columnwidth]{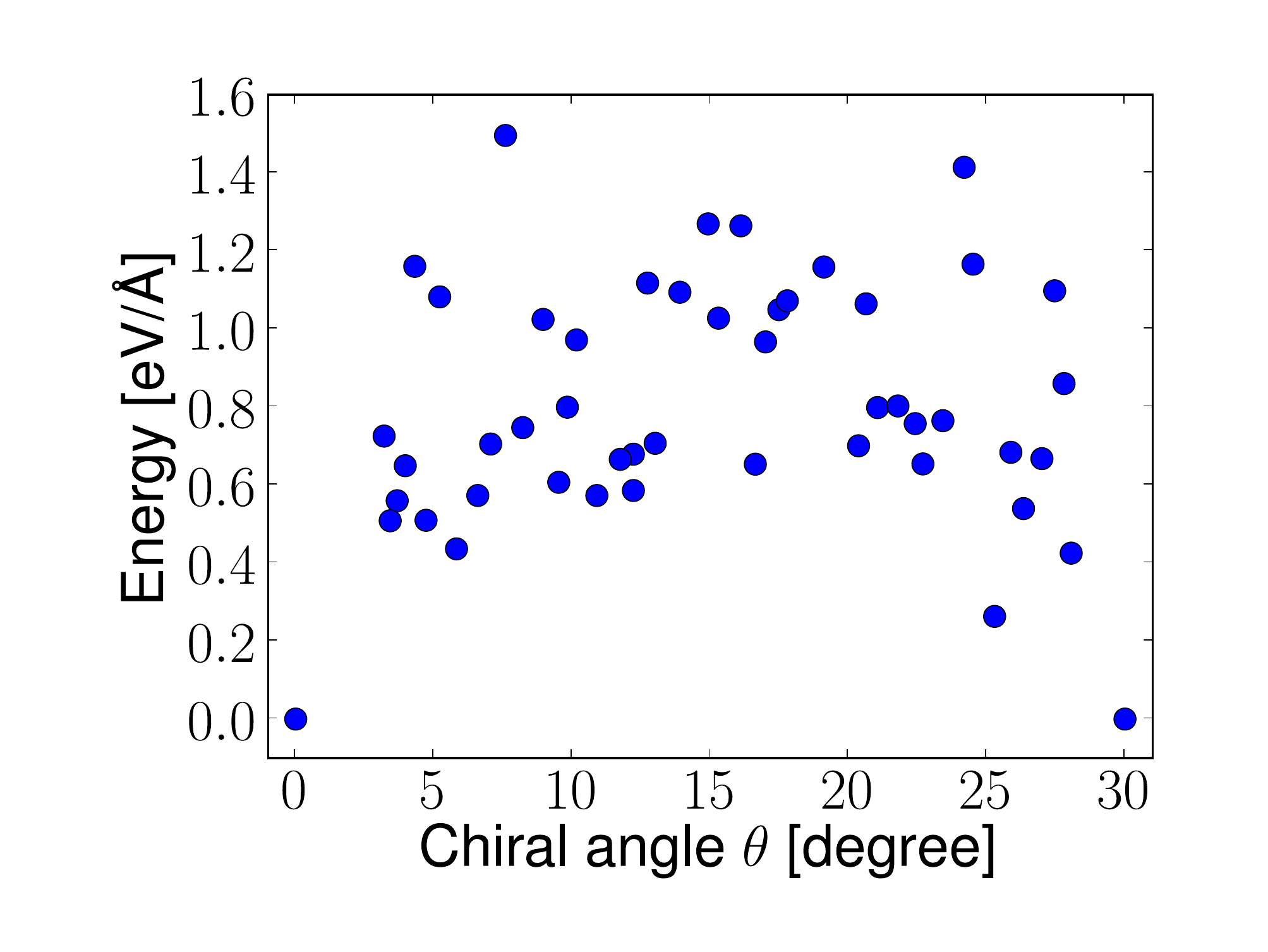}
\caption{(color online) Trends in energy density for graphene grain boundaries as a function of the chiral angle. The zero-energy means pristine graphene, and is achieved only for zigzag ($\theta=0^{\circ}$) and armchair ($\theta=30^{\circ}$) chiral angles.}
\label{fig3}
\end{figure}

The GB energy per unit length is
\begin{equation}
\varepsilon_\text{GB}=\frac{1}{L}\sum_{i \in \text{GB}} (\epsilon_i-\egr) \text{ ,}
\end{equation}
measuring how much GB costs energy relative to the same number of carbon atoms in graphene. The energies for the whole ensemble of GBs are shown in Fig.\ref{fig3}, as a function of the chiral angle. The value $\varepsilon_\text{GB}=0.0$~eV/\AA\ appears only with $\theta=0^\circ$ and $\theta=30^\circ$---meaning pristine graphene.

Most GB energies are less than or equal to the edge energies of the free graphene edges, albeit with variation. This means that GBs regain, on average, other ribbon's edge energy during the merging, and the energy of fusion ranges from $1\ldots2$~eV/\AA; on average half of the free edges' dangling bonds get passivated. The picture is not this simple, however, as MD simulation creates different polygons that cause strain. The randomness of the polygon formation is manifested by the energy variations in Fig.\ref{fig3}. Compared to ideal GBs with pentagons and heptagons only, our ensemble reveals the full complexity that rough GBs have.

\begin{figure}
\includegraphics[width=1.0\columnwidth]{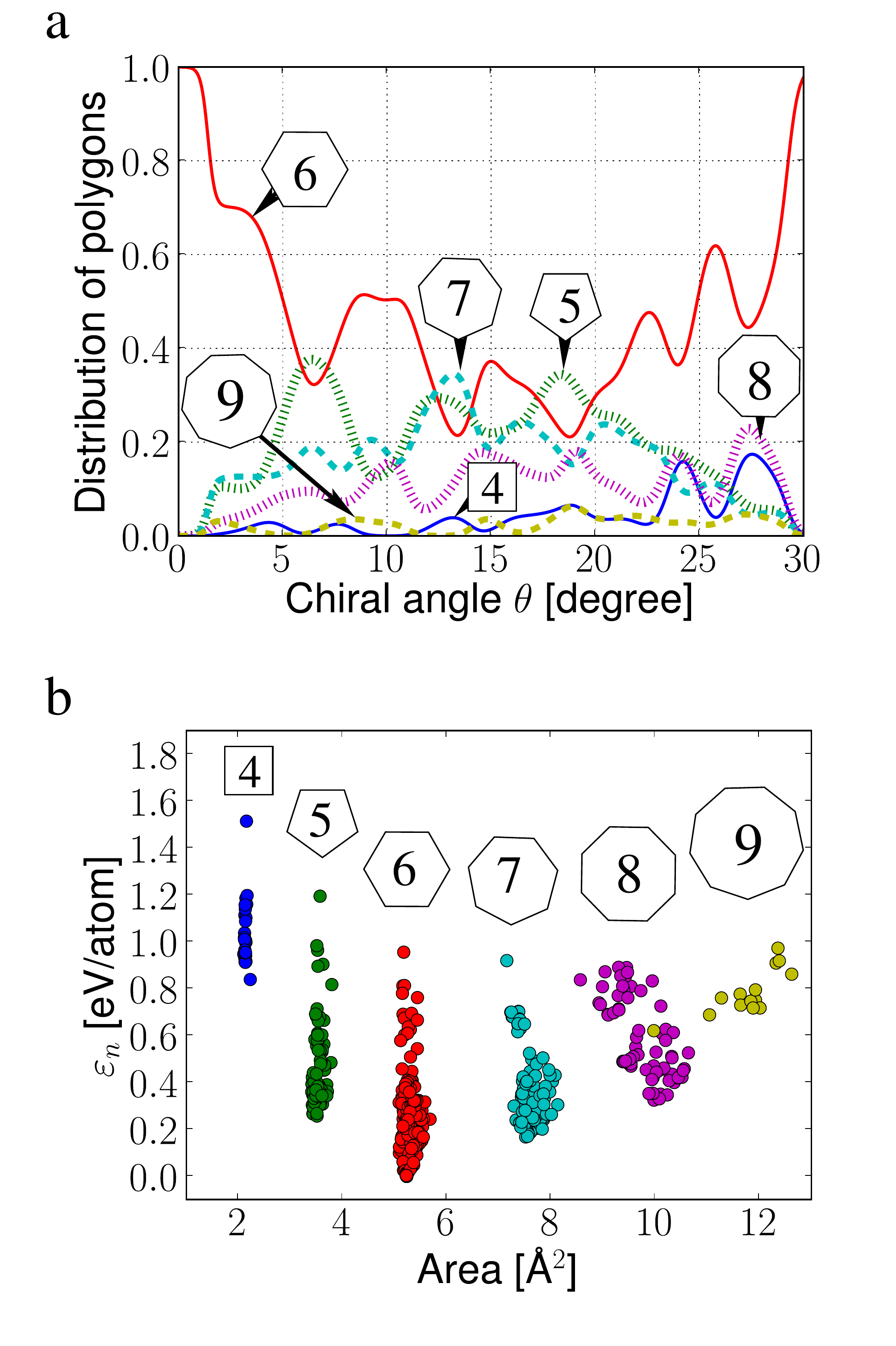}
\caption{(color online) a) Relative number of polygons in grain boundaries as a function of chiral angle. The largest lattice mismatch, near chiral angle $\sim 15^\circ$, is manifested by the appearance of mainly pentagons and heptagons, but also squares, octagons and nonagons. In b) each dot represents one polygon in all the grain boundary polygons in all of our $48$ samples. Abscissa shows the area of that polygon (determined by triangulation) and ordinate shows the polygon energy cost per atom, relative to an atom in pristine graphene.}
\label{fig4}
\end{figure}

GB energies are the highest around $\theta\sim 15^\circ$, which can be understood from structural analysis, shown in Fig.\ref{fig4}a as polygon distribution within the GB zone. What we count into GB zone are all the polygons that were \emph{not} part of the ribbons prior to merging. Polygons larger than nonagons, which appear more like vacancies instead of polygons, are omitted here. It is around $\theta\sim 15^\circ$ where the abundance of hexagons is at minimum and GBs are invaded by other polygons. The abundance of pentagons and heptagons is as high as the abundance of hexagons, but also squares, octagons and nonagons are found. Close to $\theta\sim 0^\circ$ and $\theta\sim 30^\circ$ hexagons prevail. From this we may conclude that around $\theta\sim 15^\circ$ the edge geometries have the largest mismatch, resulting in various polygons, consequent strains, and high energy.

To understand how much different polygons cost energy, we used the quantity
\begin{equation}
\varepsilon_n = \frac{1}{n}\sum_{i\in n\text{-gon}} (\epsilon_i-\egr),
\label{eq:polye}
\end{equation}
measuring how much atoms, on average, cost more in $n$-gon relative to atoms in pristine graphene; this quantity is for illustration only and considers polygons as separate items---the sum of $\varepsilon_n$'s for given GB is not the total energy. Fig.\ref{fig4}b shows $\varepsilon_n$'s for the polygons within GB zones as a function of the area of the polygon (determined by triangulation). For simple geometrical reasons for small polygons the area distribution is narrow; large polygons have more freedom to change their shape. For small polygons the distributions in $\varepsilon_n$, on the contrary, are wider; this is partly due to smaller $n$ in Eq.(\ref{eq:polye}). Clearly, the cost $\varepsilon_n$ of any polygon depends on its environment, just as it also depends for hexagons, for which $\varepsilon_6 = 0\ldots 1$~eV. But note that Fig.\ref{fig4}b already contains the effect of the polygon environment and all potential cross-correlations (such as pentagons often neighboring heptagons). It would be interesting to investigate polygon statistics also from transmission electron microscopy, now that aberration-corrected measurements can achieve atom accuracy~\cite{Meyer_NL_08}.

\section{Trends in buckling (inflection angles)}
\label{sec:buckling}

\begin{figure}[tb!]
\includegraphics[width=1.0\columnwidth]{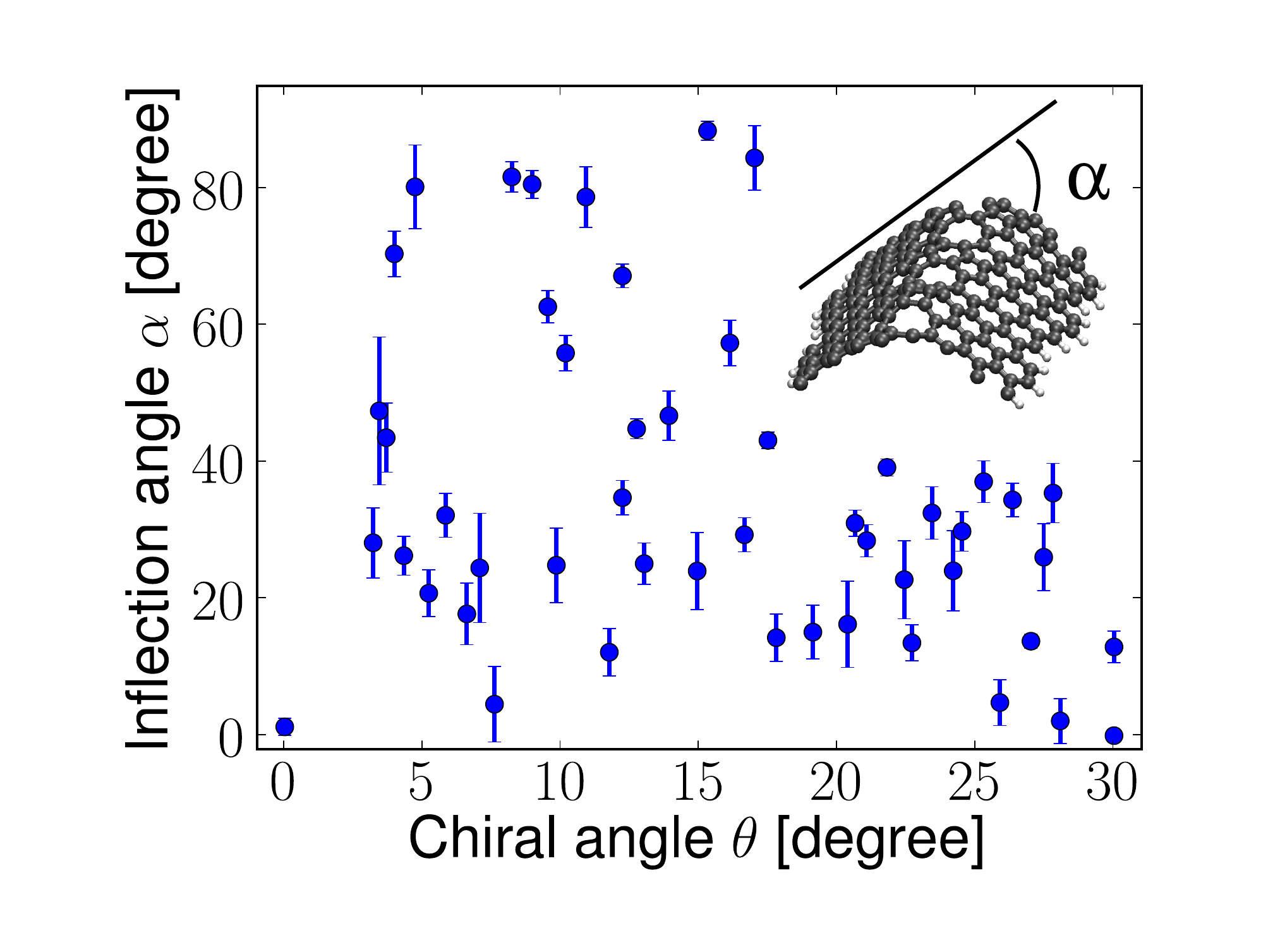}
\caption{(color online) Inflection angles after optimizations for all grain boundary samples; $0^{\circ}$ means planar structure (occurring only with pristine graphene, $\theta=0^\circ$ and $\theta=30^\circ$). The error bars remind of technical ambiguities in angle determination. The main trend here is that zigzag edges cause both large and small inflection, whereas armchair edges cause only small inflections. Inset: inflection angle illustrated.}
\label{fig5}
\end{figure}

In the generation process GBs are free to deform. The optimized GBs will typically end up having inflection angles, as the inset in Fig.\ref{fig5} illustrates. The data points in Fig.\ref{fig5} show the inflection angles for the GB samples as a function of chirality. Flat GBs occur with $\theta=0^\circ$ and $\theta=30^\circ$, that is, with pristine graphene alone. 

The notable trend in this scattered plot is the systematically small inflection angles, meaning flat GBs, around $\theta \sim 30^\circ$; near $\theta \sim 0^\circ$ inflection angles are more scattered meaning geometries that vary from flat to sharply kinked GBs. This trend can be understood by edge profiles: the stronger dangling bonds at zigzag edge, when brought into contact with another edge, can induce larger distortions than inert armchair edge (dangling bonds at armchair are partly quenched by triple-bond formation). The steps in edge profiles near $\theta\sim 0^\circ$, moreover, are $2.1$~\AA, whereas the steps in edge profiles near $\theta\sim 30^\circ$ are only $1.2$~\AA. This means that, to make bonds, the edge atoms near $\theta\sim 0^\circ$ need more pulling than edge atoms near $\theta\sim 30^\circ$, causing the buckling. 

It is clear that the numbers in Fig.\ref{fig5} have no direct relevance as such, because our geometrically optimized GB samples are in vacuum, and measure only $\sim 20$~\AA\ across the GB. We argue, however, that the inflection angle measures how much given GB would buckle in an experiment---large inflection angle meaning tendency to stick out (sticking out would be bound by geometric constraints and by surface adhesion). For example, a GB with $\theta\sim 30^\circ$ on a support will always remain smooth, whereas a GB with $\theta\sim 0^\circ$ will either remain smooth or buckle, to be seen as a ridge or a mountain range in scanning probe experiments.\cite{biedermann_PRB_09} Indeed, this buckling tendency was seen in an experiment by \v{C}ervenka \emph{at al.}, measuring corrugation heights up to $3$~\AA\ within GB regions\cite{Cervenka_naturephys_09}. Small corrugation heights and inflection angles are natural in samples on flat substrates, but also large inflection angles are realistic, for example in soot particles.~\cite{Muller_CatToday_05}

\section{Trends in chemical properties}
Now we pose the general question: how reactive are GBs? We approach this question by examining hydrogen adsorption energies and dangling bonds (DB). To this end, we first develop a connection between hydrogen adsorption energy and the electronic structure given by DFTB.

\begin{figure}
\includegraphics[width=1.0\columnwidth]{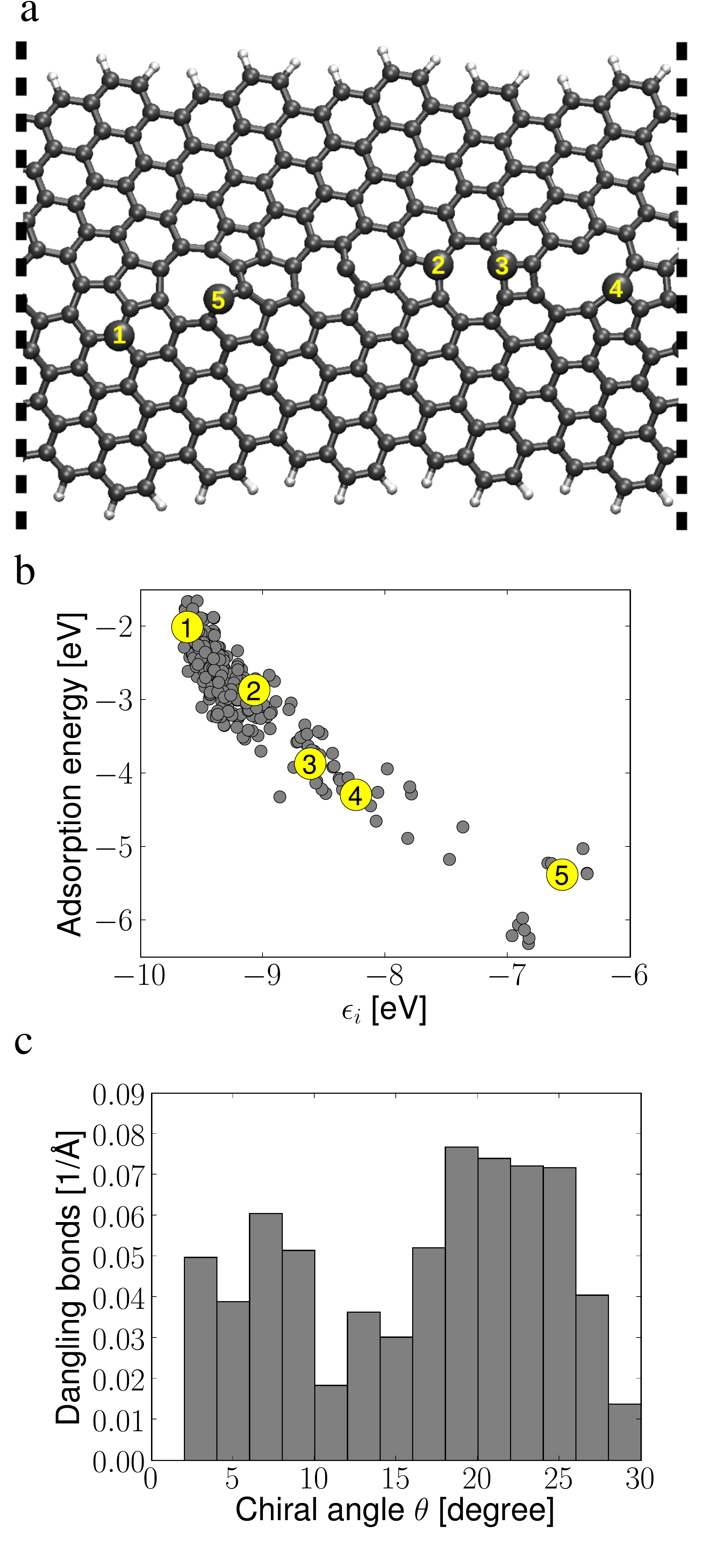}
\caption{(color online) Characterizing reactivity in grain boundaries.
a) Grain boundary sample with $(9,5)$ chirality, including various polygons. b) Characterizing chemical reactivity from the DFTB electronic structure directly: we calculated hydrogen adsorption energy for selected $350$ atoms as adsorption sites and plot for each atom adsorption energy and $\epsilon_i$. Given the correlation between the adsorption energy and $\epsilon_i$, we access reactivity directly from the electronic structure, without additional calculations. The adsorption site numbers refer to panel a); atom \ding{192}, having three $\sim 120^\circ$ angles has a small adsorption energy, atom \ding{196} has a dangling bond, and other atoms have strained environment. c) Using the correlation established in panels a) and b), we calculate the averaged density of dangling bonds per unit length along grain boundary for all samples, as a function of $\theta$.}
\label{fig:reactivity}
\end{figure}

Fig.\ref{fig:reactivity}b shows the hydrogen adsorption energies for carbon atoms in $6$ representative GB samples, from total $350$ hydrogen adsorption calculations. It appears that adsorption to given carbon atom $i$ is strong if atom's cohesion $-\epsilon_i$ decreases. Carbon atoms part of regular hexagons have adsorption around $-2$~eV (\ding{192} in Fig.\ref{fig:reactivity}a) but adsorption increases if atom is surrounded by other polygons and bonding angles deviate from $120^\circ$ (\ding{193}, \ding{194}, and \ding{195} in Fig.\ref{fig:reactivity}a). The common denominator in these examples is that the change in hydrogen adsorption energy is caused by the strain in bond angles and in bond lengths.

The strongest adsorptions around $-5 \ldots -6.5$~eV, in turn, are caused by dangling bonds that have $\epsilon_i>-7$~eV (\ding{196} in Fig.\ref{fig:reactivity}); such strong adsorption never occurs with three-coordinated atoms. (We gave the argument about DB energetics already in Sec.\ref{sec:DFTB}.) We note that DFTB hydrogen adsorption energy to zigzag ($5.8$~eV), for example, agrees reasonably with DFT energy ($5.4$~eV)\cite{koskinen_PRL_08}. Therefore, we can characterize the reactivity \emph{directly} by the DFTB electronic structure using the quantities $\epsilon_i$, without any adsorption calculations. Surely, dangling bonds can be identified from geometry by defining criteria for coordination numbers, but this approach is prone to errors, especially in disordered regions that GBs are.

Using $\epsilon_i>-7$~eV as a criterion for a dangling bond, we then analyzed the reactivity for the whole GB ensemble. Fig.\ref{fig:reactivity}c shows the average number of dangling bonds per unit length in a GB with a given chiral angle. The highest density, one DB per $\sim 14$~\AA, occurs with $\theta \sim 20^\circ$. The lowest density, one DB per $\sim 30$~\AA, occurs with $\theta \sim 15^\circ$; pristine graphene with no DBs becomes probable with $\theta\sim 0^\circ$ and $\theta \sim 30^\circ$. 

By analyzing the histogram in Fig.\ref{fig:reactivity}c another way, $30$~\%\ of the samples have no DBs, $54$~\%\ of the samples have one DB every $7.5 \ldots 20$~\AA, and $16$~\%\ of the samples have one DB every $20 \ldots 50$~\AA. These numbers agree with a recent scanning tunneling microscopy (STM) experiment. Namely, dangling bonds are highly localized states just below the Fermi-level, and hence seen as a bump in constant-current STM with low bias. The STM images of \v{C}ervenka \emph{et al.} in Ref.~\onlinecite{Cervenka_naturephys_09} show periodic appearance of sharp peaks, with $5\ldots 20$~\AA\ periodicity for $53$~\%\ of the samples and with $\gtrsim 20$~\AA\ periodicity for $47$~\% of the samples. Even if these periodicities should be caused by adsorbed impurities and not from bonds that dangle, it is still likely to be result from \emph{reactive} sites within GBs---and answers the original question we posed in this section. The agreement with experiment gives confidence in the objectivity of the GB construction process. Understanding the trends in defects and reactivity will hopefully help in the design of functionalized graphene compounds~\cite{Geim_Science_09}.

\section{Trends in vibrational properties}
The structure of a GB, given its constituent polygons, affects directly on its vibrational spectrum, and gives experimentally complementary information.

\begin{figure}
\includegraphics[width=1.0\columnwidth]{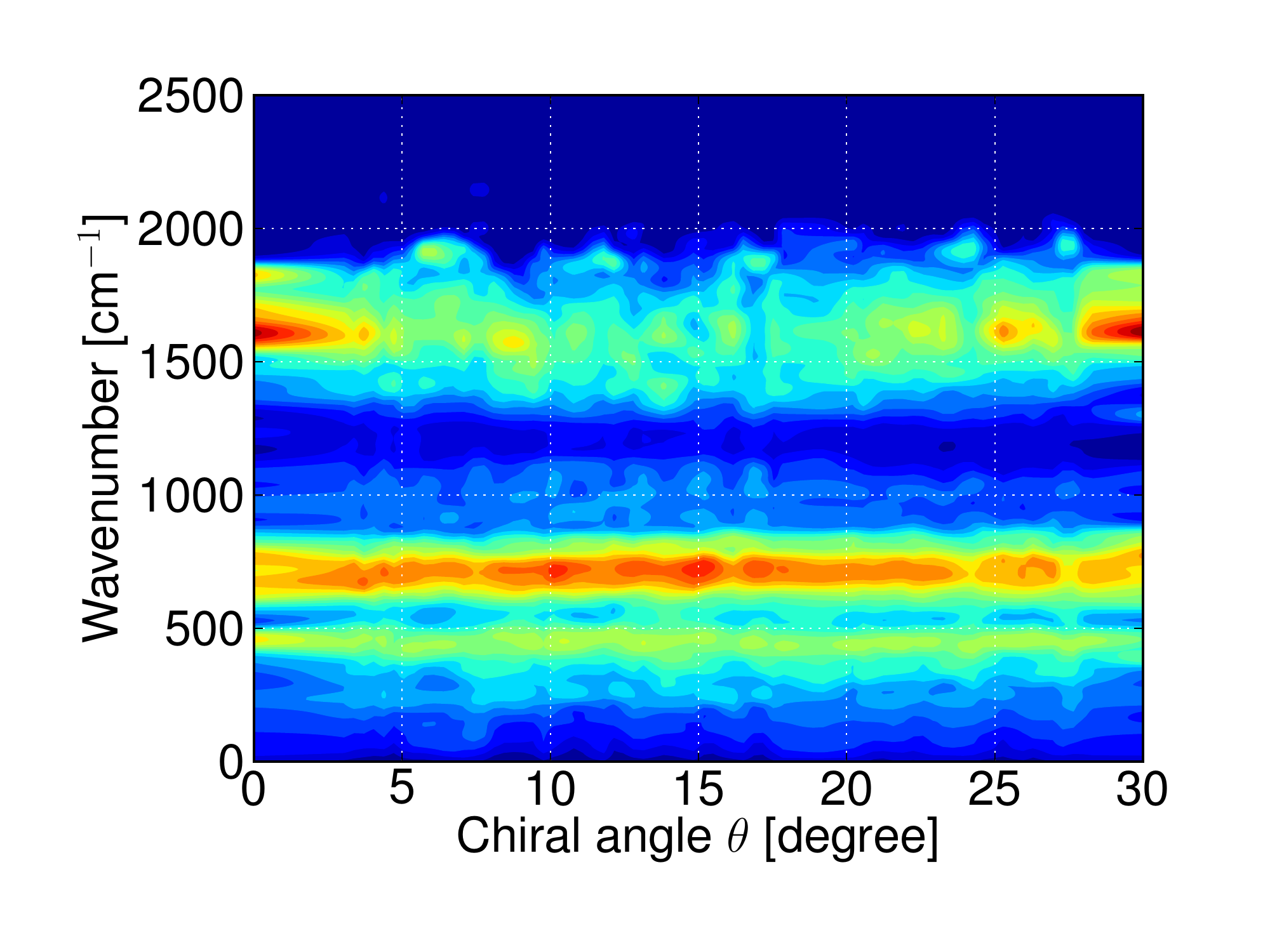}
\caption{(color online) Projected vibrational density of states (PVDOS) as a function of the wavenumber and the chiral angle. PVDOS for each sample (given $\theta$) was calculated by solving vibrational spectrum from the dynamical matrix, and projecting the vibration eigenmodes on atoms within the grain boundary; the contour plot is then gathered and smoothed from all the $48$ samples. Intensity increases from blue to red.}
\label{fig:vibr}
\end{figure}

Fig.\ref{fig:vibr} shows the projected vibrational density of states (PVDOS) for the GBs as a function of chiral angle and wave number. PVDOS was calculated by solving vibrations for the whole GB, by projecting the eigenmodes to atoms within GB area, and by renormalizing the modes---for all the $48$ GB samples. Hence Fig.\ref{fig:vibr} shows a lot of data in a compact form.

Spectra show three main bands, two at low energies $\sim 750$~cm$^{-1}$ and $\sim 500$~cm$^{-1}$, and one---the so-called G-band in Raman spectroscopy---at high energy $\sim1600$~cm$^{-1}$. The band at $\sim 500$~cm$^{-1}$ is steady across all $\theta$'s and can not be used for structural identification. The G-band, in turn, loses intensity and comes down some $100$~cm$^{-1}$ in energy around $\theta\sim 15^\circ$. This can be explained by structure. As discussed in Sec.\ref{sec:structure}, around $\theta\sim 15^\circ$ hexagons are at relative minimum, implying a non-uniform and less rigid structure, and causing floppiness in high-energy modes. The high-energy modes have bond stretching between nearest-neighbors and are hence sensitive to local structural changes. Low-energy modes, again, are more collective and hence insensitive to local changes in structure, as long as they remain somehow ``graphitic''. If Fig.\ref{fig:vibr} and Fig.\ref{fig4} are compared carefully, one finds that the local intensity maxima of the G-band occur precisely for $\theta$ with local maxima in the abundance of hexagons. The intensity increase around $\sim750$~cm$^{-1}$ and $\theta\sim 15^\circ$ is mainly due to renormalization of PVDOS.

The observations above are qualitatively supported by earlier experiments. Using Raman spectroscopy, a G-band shift from $1600$~cm$^{-1}$ to $1510$~cm$^{-1}$ was reported by Ferrari and Robertson in Ref.~\onlinecite{Ferrari_PRB_00}; the shift was identified to be a consequence of structural change from nanocrystalline graphite to amorphous phase. While still being mainly planar (apart from inflection), GB zone around $\theta\sim 15^\circ$, such as Fig.\ref{fig:reactivity}a with $\theta=20.6^\circ$, indeed appears amorphous.

\section{Concluding remarks}
Some defects, like singular Stone-Wales defect or reczag edge of graphene, have their own name because they are well defined. Grain boundaries, on the contrary, are like snowflakes---there is no flake like another, but it's enough to know that they are roughly hexagonal, flat, small and cold. For the same reason it's enough to know how grain boundaries usually look and feel. Knowing trends is valuable.

We investigated GBs from different viewpoints, discovering trends with complementary information, measurable also experimentally. We investigated (i) geometry (polygons and inflection angles) measurable with transmission electron microscope or atomic probe microscope; (ii) energy, that is manifested in geometry and thereby measurable; (iii) reactivity, measurable with adsorption experiments; (iv) vibrational properties, measurable with Raman spectroscopy. We are confident that the trends are genuine, part because of agreements with earlier experiments, part because the trends all make intuitive sense: energy, inflection angles, reactivity and vibration trends make sense given the structure, and the structure trends make sense given the graphene edge mismatch. We hope the trends help to explain experiments---and also help simulations and experiments to design graphene structures for given functions.

\section*{Acknowledgments}
This work was supported by the Academy of Finland (project $121701$ the and the FINNANO MEP consortium) and Finnish Cultural Foundation. Computational resources were provided by the Nanoscience Center in the University of Jyv\"askyl\"a and by the Finnish IT Center for Science (CSC) in Espoo.


\begin{thebibliography}{35}
\expandafter\ifx\csname natexlab\endcsname\relax\def\natexlab#1{#1}\fi
\expandafter\ifx\csname bibnamefont\endcsname\relax
  \def\bibnamefont#1{#1}\fi
\expandafter\ifx\csname bibfnamefont\endcsname\relax
  \def\bibfnamefont#1{#1}\fi
\expandafter\ifx\csname citenamefont\endcsname\relax
  \def\citenamefont#1{#1}\fi
\expandafter\ifx\csname url\endcsname\relax
  \def\url#1{\texttt{#1}}\fi
\expandafter\ifx\csname urlprefix\endcsname\relax\def\urlprefix{URL }\fi
\providecommand{\bibinfo}[2]{#2}
\providecommand{\eprint}[2][]{\url{#2}}

\bibitem[{\citenamefont{\v{C}ervenka and Flipse}(2007)}]{Cervenka_JPCS_07}
\bibinfo{author}{\bibfnamefont{J.}~\bibnamefont{\v{C}ervenka}}
  \bibnamefont{and} \bibinfo{author}{\bibfnamefont{C.}~\bibnamefont{Flipse}},
  \bibinfo{journal}{J. Phys. Conf. Ser.} \textbf{\bibinfo{volume}{61}},
  \bibinfo{pages}{190} (\bibinfo{year}{2007}).

\bibitem[{\citenamefont{\v{C}ervenka et~al.}(2009)\citenamefont{\v{C}ervenka,
  Katsnelson, and Flipse}}]{Cervenka_naturephys_09}
\bibinfo{author}{\bibfnamefont{J.}~\bibnamefont{\v{C}ervenka}},
  \bibinfo{author}{\bibfnamefont{M.}~\bibnamefont{Katsnelson}},
  \bibnamefont{and} \bibinfo{author}{\bibfnamefont{C.}~\bibnamefont{Flipse}},
  \bibinfo{journal}{Nature Physics} \textbf{\bibinfo{volume}{5}},
  \bibinfo{pages}{840} (\bibinfo{year}{2009}).

\bibitem[{\citenamefont{\v{C}ervenka and Flipse}(2009)}]{cervenka_PRB_09}
\bibinfo{author}{\bibfnamefont{J.}~\bibnamefont{\v{C}ervenka}}
  \bibnamefont{and} \bibinfo{author}{\bibfnamefont{C.~F.~J.}
  \bibnamefont{Flipse}}, \bibinfo{journal}{Phys. Rev. B}
  \textbf{\bibinfo{volume}{79}}, \bibinfo{pages}{195429}
  (\bibinfo{year}{2009}).

\bibitem[{\citenamefont{Iijima et~al.}(1996)\citenamefont{Iijima, Wakabayashi,
  and Achiba}}]{Iijima_JPC_96}
\bibinfo{author}{\bibfnamefont{S.}~\bibnamefont{Iijima}},
  \bibinfo{author}{\bibfnamefont{T.}~\bibnamefont{Wakabayashi}},
  \bibnamefont{and} \bibinfo{author}{\bibfnamefont{Y.}~\bibnamefont{Achiba}},
  \bibinfo{journal}{J. Phys. Chem.} \textbf{\bibinfo{volume}{100}},
  \bibinfo{pages}{5839} (\bibinfo{year}{1996}).

\bibitem[{\citenamefont{Boehman et~al.}(2005)\citenamefont{Boehman, Song, and
  Alam}}]{Boehman_EF_05}
\bibinfo{author}{\bibfnamefont{A.}~\bibnamefont{Boehman}},
  \bibinfo{author}{\bibfnamefont{J.}~\bibnamefont{Song}}, \bibnamefont{and}
  \bibinfo{author}{\bibfnamefont{M.}~\bibnamefont{Alam}},
  \bibinfo{journal}{Energy \& Fuels} \textbf{\bibinfo{volume}{19}},
  \bibinfo{pages}{1857} (\bibinfo{year}{2005}).

\bibitem[{\citenamefont{M\"uller et~al.}(2005)\citenamefont{M\"uller, Su,
  Jentoft, Kr\"ohnert, Jentoft, and Schl\"ogl}}]{Muller_CatToday_05}
\bibinfo{author}{\bibfnamefont{J.-O.} \bibnamefont{M\"uller}},
  \bibinfo{author}{\bibfnamefont{D.}~\bibnamefont{Su}},
  \bibinfo{author}{\bibfnamefont{R.}~\bibnamefont{Jentoft}},
  \bibinfo{author}{\bibfnamefont{J.}~\bibnamefont{Kr\"ohnert}},
  \bibinfo{author}{\bibfnamefont{F.}~\bibnamefont{Jentoft}}, \bibnamefont{and}
  \bibinfo{author}{\bibfnamefont{R.}~\bibnamefont{Schl\"ogl}},
  \bibinfo{journal}{Catalysis Today} \textbf{\bibinfo{volume}{102-103}},
  \bibinfo{pages}{259} (\bibinfo{year}{2005}).

\bibitem[{\citenamefont{Varchon et~al.}(2008)\citenamefont{Varchon, Mallet,
  Magaud, and Veuillen}}]{varchon_PRB_08}
\bibinfo{author}{\bibfnamefont{F.}~\bibnamefont{Varchon}},
  \bibinfo{author}{\bibfnamefont{P.}~\bibnamefont{Mallet}},
  \bibinfo{author}{\bibfnamefont{L.}~\bibnamefont{Magaud}}, \bibnamefont{and}
  \bibinfo{author}{\bibfnamefont{J.-Y.} \bibnamefont{Veuillen}},
  \bibinfo{journal}{Phys. Rev. B} \textbf{\bibinfo{volume}{77}},
  \bibinfo{pages}{165415} (\bibinfo{year}{2008}).

\bibitem[{\citenamefont{Chae et~al.}(2009)\citenamefont{Chae, G\"unes, Kim,
  Kim, Han, Kim, Shin, Yoon, Choi, Park et~al.}}]{Chae_AdvMat_09}
\bibinfo{author}{\bibfnamefont{S.}~\bibnamefont{Chae}},
  \bibinfo{author}{\bibfnamefont{F.}~\bibnamefont{G\"unes}},
  \bibinfo{author}{\bibfnamefont{K.}~\bibnamefont{Kim}},
  \bibinfo{author}{\bibfnamefont{E.}~\bibnamefont{Kim}},
  \bibinfo{author}{\bibfnamefont{G.}~\bibnamefont{Han}},
  \bibinfo{author}{\bibfnamefont{S.}~\bibnamefont{Kim}},
  \bibinfo{author}{\bibfnamefont{H.-J.} \bibnamefont{Shin}},
  \bibinfo{author}{\bibfnamefont{S.-M.} \bibnamefont{Yoon}},
  \bibinfo{author}{\bibfnamefont{J.-Y.} \bibnamefont{Choi}},
  \bibinfo{author}{\bibfnamefont{M.}~\bibnamefont{Park}}, \bibnamefont{et~al.},
  \bibinfo{journal}{Adv. Mater.} \textbf{\bibinfo{volume}{21}},
  \bibinfo{pages}{2328} (\bibinfo{year}{2009}).

\bibitem[{\citenamefont{Terrones et~al.}(2002)\citenamefont{Terrones, Terrones,
  and Terrones}}]{Terrones_SC_02}
\bibinfo{author}{\bibfnamefont{M.}~\bibnamefont{Terrones}},
  \bibinfo{author}{\bibfnamefont{G.}~\bibnamefont{Terrones}}, \bibnamefont{and}
  \bibinfo{author}{\bibfnamefont{H.}~\bibnamefont{Terrones}},
  \bibinfo{journal}{Structural Chemistry} \textbf{\bibinfo{volume}{13}},
  \bibinfo{pages}{373} (\bibinfo{year}{2002}).

\bibitem[{\citenamefont{Lau et~al.}(2007)\citenamefont{Lau, McCulloch, Marks,
  Madsen, and Rode}}]{Lau_PRB_07}
\bibinfo{author}{\bibfnamefont{D.}~\bibnamefont{Lau}},
  \bibinfo{author}{\bibfnamefont{D.}~\bibnamefont{McCulloch}},
  \bibinfo{author}{\bibfnamefont{N.}~\bibnamefont{Marks}},
  \bibinfo{author}{\bibfnamefont{N.}~\bibnamefont{Madsen}}, \bibnamefont{and}
  \bibinfo{author}{\bibfnamefont{A.}~\bibnamefont{Rode}},
  \bibinfo{journal}{Phys. Rev. B} \textbf{\bibinfo{volume}{75}},
  \bibinfo{pages}{233408} (\bibinfo{year}{2007}).

\bibitem[{\citenamefont{Charlier}(2002)}]{Charlier_AccChemRes_02}
\bibinfo{author}{\bibfnamefont{J.-C.} \bibnamefont{Charlier}},
  \bibinfo{journal}{Phys. Rev. B} \textbf{\bibinfo{volume}{35}},
  \bibinfo{pages}{1063} (\bibinfo{year}{2002}).

\bibitem[{\citenamefont{Ren et~al.}(1999)\citenamefont{Ren, Huang, Wang, Wen,
  Xu, Wang, Calvet, Chen, Klemic, and Reed}}]{Ren_APL_99}
\bibinfo{author}{\bibfnamefont{Z.}~\bibnamefont{Ren}},
  \bibinfo{author}{\bibfnamefont{Z.}~\bibnamefont{Huang}},
  \bibinfo{author}{\bibfnamefont{D.}~\bibnamefont{Wang}},
  \bibinfo{author}{\bibfnamefont{J.}~\bibnamefont{Wen}},
  \bibinfo{author}{\bibfnamefont{J.}~\bibnamefont{Xu}},
  \bibinfo{author}{\bibfnamefont{J.}~\bibnamefont{Wang}},
  \bibinfo{author}{\bibfnamefont{L.}~\bibnamefont{Calvet}},
  \bibinfo{author}{\bibfnamefont{J.}~\bibnamefont{Chen}},
  \bibinfo{author}{\bibfnamefont{J.}~\bibnamefont{Klemic}}, \bibnamefont{and}
  \bibinfo{author}{\bibfnamefont{M.}~\bibnamefont{Reed}},
  \bibinfo{journal}{Appl. Phys. Lett.} \textbf{\bibinfo{volume}{75}},
  \bibinfo{pages}{1086} (\bibinfo{year}{1999}).

\bibitem[{\citenamefont{Geim}(2009)}]{Geim_Science_09}
\bibinfo{author}{\bibfnamefont{A.}~\bibnamefont{Geim}},
  \bibinfo{journal}{Science} \textbf{\bibinfo{volume}{324}},
  \bibinfo{pages}{1530} (\bibinfo{year}{2009}).

\bibitem[{\citenamefont{Stankovich et~al.}(2006)\citenamefont{Stankovich,
  Dikin, Dommet, Kohlhaas, Zimney, Stach, Piner, Nguyen, and
  Ruoff}}]{Stankovich_Nature_06}
\bibinfo{author}{\bibfnamefont{S.}~\bibnamefont{Stankovich}},
  \bibinfo{author}{\bibfnamefont{D.~A.} \bibnamefont{Dikin}},
  \bibinfo{author}{\bibfnamefont{G.~H.~B.} \bibnamefont{Dommet}},
  \bibinfo{author}{\bibfnamefont{K.~M.} \bibnamefont{Kohlhaas}},
  \bibinfo{author}{\bibfnamefont{E.~J.} \bibnamefont{Zimney}},
  \bibinfo{author}{\bibfnamefont{E.~A.} \bibnamefont{Stach}},
  \bibinfo{author}{\bibfnamefont{R.~D.} \bibnamefont{Piner}},
  \bibinfo{author}{\bibfnamefont{S.~T.} \bibnamefont{Nguyen}},
  \bibnamefont{and} \bibinfo{author}{\bibfnamefont{R.~S.} \bibnamefont{Ruoff}},
  \bibinfo{journal}{Nature} \textbf{\bibinfo{volume}{442}},
  \bibinfo{pages}{282} (\bibinfo{year}{2006}).

\bibitem[{\citenamefont{{Castro Neto} et~al.}(2009)\citenamefont{{Castro Neto},
  Guinea, and Peres}}]{Neto_RevModPhys_09}
\bibinfo{author}{\bibfnamefont{A.}~\bibnamefont{{Castro Neto}}},
  \bibinfo{author}{\bibfnamefont{F.}~\bibnamefont{Guinea}}, \bibnamefont{and}
  \bibinfo{author}{\bibfnamefont{N.}~\bibnamefont{Peres}},
  \bibinfo{journal}{Rev. Mod. Phys.} \textbf{\bibinfo{volume}{81}},
  \bibinfo{pages}{109} (\bibinfo{year}{2009}).

\bibitem[{\citenamefont{Biedermann et~al.}(2009)\citenamefont{Biedermann,
  Bolen, Capano, Zemlyanov, and Reifenberger}}]{biedermann_PRB_09}
\bibinfo{author}{\bibfnamefont{L.~B.} \bibnamefont{Biedermann}},
  \bibinfo{author}{\bibfnamefont{M.~L.} \bibnamefont{Bolen}},
  \bibinfo{author}{\bibfnamefont{M.~A.} \bibnamefont{Capano}},
  \bibinfo{author}{\bibfnamefont{D.}~\bibnamefont{Zemlyanov}},
  \bibnamefont{and} \bibinfo{author}{\bibfnamefont{R.~G.}
  \bibnamefont{Reifenberger}}, \bibinfo{journal}{Phys. Rev. B}
  \textbf{\bibinfo{volume}{79}}, \bibinfo{pages}{125411}
  (\bibinfo{year}{2009}).

\bibitem[{\citenamefont{Gan et~al.}(2003)\citenamefont{Gan, Chu, and
  Qiao}}]{gan_SS_03}
\bibinfo{author}{\bibfnamefont{Y.}~\bibnamefont{Gan}},
  \bibinfo{author}{\bibfnamefont{W.}~\bibnamefont{Chu}}, \bibnamefont{and}
  \bibinfo{author}{\bibfnamefont{L.}~\bibnamefont{Qiao}},
  \bibinfo{journal}{Surface Science} \textbf{\bibinfo{volume}{539}},
  \bibinfo{pages}{120} (\bibinfo{year}{2003}).

\bibitem[{\citenamefont{Simonis et~al.}(2002)\citenamefont{Simonis, Goffaux,
  Thiry, Biro, Lambin, and Meunier}}]{simonis_SS_02}
\bibinfo{author}{\bibfnamefont{P.}~\bibnamefont{Simonis}},
  \bibinfo{author}{\bibfnamefont{C.}~\bibnamefont{Goffaux}},
  \bibinfo{author}{\bibfnamefont{P.~A.} \bibnamefont{Thiry}},
  \bibinfo{author}{\bibfnamefont{L.~P.} \bibnamefont{Biro}},
  \bibinfo{author}{\bibfnamefont{P.}~\bibnamefont{Lambin}}, \bibnamefont{and}
  \bibinfo{author}{\bibfnamefont{V.}~\bibnamefont{Meunier}},
  \bibinfo{journal}{Surface Science} \textbf{\bibinfo{volume}{511}},
  \bibinfo{pages}{319} (\bibinfo{year}{2002}).

\bibitem[{\citenamefont{Yao et~al.}(1999)\citenamefont{Yao, Postma, Balents,
  and Dekker}}]{yao_nature_99}
\bibinfo{author}{\bibfnamefont{Z.}~\bibnamefont{Yao}},
  \bibinfo{author}{\bibfnamefont{H.~W.~C.} \bibnamefont{Postma}},
  \bibinfo{author}{\bibfnamefont{L.}~\bibnamefont{Balents}}, \bibnamefont{and}
  \bibinfo{author}{\bibfnamefont{C.}~\bibnamefont{Dekker}},
  \bibinfo{journal}{Nature} \textbf{\bibinfo{volume}{402}},
  \bibinfo{pages}{273} (\bibinfo{year}{1999}).

\bibitem[{\citenamefont{Oyang et~al.}(2001)\citenamefont{Oyang, Huang, Cheung,
  and Lieber}}]{ouyang_science_01}
\bibinfo{author}{\bibfnamefont{M.}~\bibnamefont{Oyang}},
  \bibinfo{author}{\bibfnamefont{J.-L.} \bibnamefont{Huang}},
  \bibinfo{author}{\bibfnamefont{C.~L.} \bibnamefont{Cheung}},
  \bibnamefont{and} \bibinfo{author}{\bibfnamefont{C.~M.}
  \bibnamefont{Lieber}}, \bibinfo{journal}{Science}
  \textbf{\bibinfo{volume}{291}}, \bibinfo{pages}{97} (\bibinfo{year}{2001}).

\bibitem[{\citenamefont{Chico et~al.}(1996)\citenamefont{Chico, Crespi,
  Benedict, Louie, and Cohen}}]{chico_PRL_96}
\bibinfo{author}{\bibfnamefont{L.}~\bibnamefont{Chico}},
  \bibinfo{author}{\bibfnamefont{V.~H.} \bibnamefont{Crespi}},
  \bibinfo{author}{\bibfnamefont{L.~X.} \bibnamefont{Benedict}},
  \bibinfo{author}{\bibfnamefont{S.~G.} \bibnamefont{Louie}}, \bibnamefont{and}
  \bibinfo{author}{\bibfnamefont{M.~L.} \bibnamefont{Cohen}},
  \bibinfo{journal}{Phys. Rev. Lett.} \textbf{\bibinfo{volume}{76}},
  \bibinfo{pages}{971} (\bibinfo{year}{1996}).

\bibitem[{\citenamefont{Gu et~al.}(2007)\citenamefont{Gu, Nie, Feenstra,
  Devaty, Choyke, Chan, and Kane}}]{gu_APL_07}
\bibinfo{author}{\bibfnamefont{G.}~\bibnamefont{Gu}},
  \bibinfo{author}{\bibfnamefont{S.}~\bibnamefont{Nie}},
  \bibinfo{author}{\bibfnamefont{R.~M.} \bibnamefont{Feenstra}},
  \bibinfo{author}{\bibfnamefont{R.~P.} \bibnamefont{Devaty}},
  \bibinfo{author}{\bibfnamefont{W.~J.} \bibnamefont{Choyke}},
  \bibinfo{author}{\bibfnamefont{W.~K.} \bibnamefont{Chan}}, \bibnamefont{and}
  \bibinfo{author}{\bibfnamefont{M.~G.} \bibnamefont{Kane}},
  \bibinfo{journal}{Appl. Phys. Lett.} \textbf{\bibinfo{volume}{90}},
  \bibinfo{pages}{253507} (\bibinfo{year}{2007}).

\bibitem[{\citenamefont{{da Silva Ara\'{u}jo} and Nunes}(2010)}]{araujo_PRB_10}
\bibinfo{author}{\bibfnamefont{J.}~\bibnamefont{{da Silva Ara\'{u}jo}}}
  \bibnamefont{and} \bibinfo{author}{\bibfnamefont{R.~W.} \bibnamefont{Nunes}},
  \bibinfo{journal}{Phys. Rev. B} \textbf{\bibinfo{volume}{81}},
  \bibinfo{pages}{073408} (\bibinfo{year}{2010}).

\bibitem[{\citenamefont{Porezag et~al.}(1995)\citenamefont{Porezag, Frauenheim,
  K\"ohler, Seifert, and Kaschner}}]{porezag_PRB_95}
\bibinfo{author}{\bibfnamefont{D.}~\bibnamefont{Porezag}},
  \bibinfo{author}{\bibfnamefont{T.}~\bibnamefont{Frauenheim}},
  \bibinfo{author}{\bibfnamefont{T.}~\bibnamefont{K\"ohler}},
  \bibinfo{author}{\bibfnamefont{G.}~\bibnamefont{Seifert}}, \bibnamefont{and}
  \bibinfo{author}{\bibfnamefont{R.}~\bibnamefont{Kaschner}},
  \bibinfo{journal}{Phys. Rev. B} \textbf{\bibinfo{volume}{51}},
  \bibinfo{pages}{12947} (\bibinfo{year}{1995}).

\bibitem[{\citenamefont{Elstner et~al.}(1998)\citenamefont{Elstner, Porezag,
  Jungnickel, Elstner, Haugk, Frauenheim, Suhai, and Seifert}}]{elstner_PRB_98}
\bibinfo{author}{\bibfnamefont{M.}~\bibnamefont{Elstner}},
  \bibinfo{author}{\bibfnamefont{D.}~\bibnamefont{Porezag}},
  \bibinfo{author}{\bibfnamefont{G.}~\bibnamefont{Jungnickel}},
  \bibinfo{author}{\bibfnamefont{J.}~\bibnamefont{Elstner}},
  \bibinfo{author}{\bibfnamefont{M.}~\bibnamefont{Haugk}},
  \bibinfo{author}{\bibfnamefont{T.}~\bibnamefont{Frauenheim}},
  \bibinfo{author}{\bibfnamefont{S.}~\bibnamefont{Suhai}}, \bibnamefont{and}
  \bibinfo{author}{\bibfnamefont{G.}~\bibnamefont{Seifert}},
  \bibinfo{journal}{Phys. Rev. B} \textbf{\bibinfo{volume}{58}},
  \bibinfo{pages}{7260} (\bibinfo{year}{1998}).

\bibitem[{\citenamefont{Koskinen and M\"akinen}(2009)}]{koskinen_CMS_09}
\bibinfo{author}{\bibfnamefont{P.}~\bibnamefont{Koskinen}} \bibnamefont{and}
  \bibinfo{author}{\bibfnamefont{V.}~\bibnamefont{M\"akinen}},
  \bibinfo{journal}{Computational Materials Science}
  \textbf{\bibinfo{volume}{47}}, \bibinfo{pages}{237} (\bibinfo{year}{2009}).

\bibitem[{hot()}]{hotbit_wiki}
\bibinfo{note}{Hotbit wiki \texttt{https://trac.cc.jyu.fi/projects/hotbit}}.

\bibitem[{\citenamefont{Malola et~al.}(2008{\natexlab{a}})\citenamefont{Malola,
  H{\"a}kkinen, and Koskinen}}]{malola_PRB_08}
\bibinfo{author}{\bibfnamefont{S.}~\bibnamefont{Malola}},
  \bibinfo{author}{\bibfnamefont{H.}~\bibnamefont{H{\"a}kkinen}},
  \bibnamefont{and} \bibinfo{author}{\bibfnamefont{P.}~\bibnamefont{Koskinen}},
  \bibinfo{journal}{Phys. Rev. B} \textbf{\bibinfo{volume}{77}},
  \bibinfo{pages}{155412} (\bibinfo{year}{2008}{\natexlab{a}}).

\bibitem[{\citenamefont{Malola et~al.}(2008{\natexlab{b}})\citenamefont{Malola,
  H\"akkinen, and Koskinen}}]{malola_PRB_08b}
\bibinfo{author}{\bibfnamefont{S.}~\bibnamefont{Malola}},
  \bibinfo{author}{\bibfnamefont{H.}~\bibnamefont{H\"akkinen}},
  \bibnamefont{and} \bibinfo{author}{\bibfnamefont{P.}~\bibnamefont{Koskinen}},
  \bibinfo{journal}{Phys. Rev. B} \textbf{\bibinfo{volume}{78}},
  \bibinfo{pages}{153409} (\bibinfo{year}{2008}{\natexlab{b}}).

\bibitem[{\citenamefont{Malola et~al.}(2009)\citenamefont{Malola, H{\"a}kkinen,
  and Koskinen}}]{malola_EPJD_09}
\bibinfo{author}{\bibfnamefont{S.}~\bibnamefont{Malola}},
  \bibinfo{author}{\bibfnamefont{H.}~\bibnamefont{H{\"a}kkinen}},
  \bibnamefont{and} \bibinfo{author}{\bibfnamefont{P.}~\bibnamefont{Koskinen}},
  \bibinfo{journal}{Eur. Phys. J. D} \textbf{\bibinfo{volume}{52}},
  \bibinfo{pages}{71} (\bibinfo{year}{2009}).

\bibitem[{\citenamefont{Bitzek et~al.}(2006)\citenamefont{Bitzek, Koskinen,
  G\"ahler, Moseler, and Gumbsch}}]{bitzek_PRL_06}
\bibinfo{author}{\bibfnamefont{E.}~\bibnamefont{Bitzek}},
  \bibinfo{author}{\bibfnamefont{P.}~\bibnamefont{Koskinen}},
  \bibinfo{author}{\bibfnamefont{F.}~\bibnamefont{G\"ahler}},
  \bibinfo{author}{\bibfnamefont{M.}~\bibnamefont{Moseler}}, \bibnamefont{and}
  \bibinfo{author}{\bibfnamefont{P.}~\bibnamefont{Gumbsch}},
  \bibinfo{journal}{Phys. Rev. Lett.} \textbf{\bibinfo{volume}{97}},
  \bibinfo{pages}{170201} (\bibinfo{year}{2006}).

\bibitem[{\citenamefont{Koskinen et~al.}(2008)\citenamefont{Koskinen, Malola,
  and H\"akkinen}}]{koskinen_PRL_08}
\bibinfo{author}{\bibfnamefont{P.}~\bibnamefont{Koskinen}},
  \bibinfo{author}{\bibfnamefont{S.}~\bibnamefont{Malola}}, \bibnamefont{and}
  \bibinfo{author}{\bibfnamefont{H.}~\bibnamefont{H\"akkinen}},
  \bibinfo{journal}{Phys. Rev. Lett.} \textbf{\bibinfo{volume}{101}},
  \bibinfo{pages}{115502} (\bibinfo{year}{2008}).

\bibitem[{\citenamefont{Koskinen et~al.}(2009)\citenamefont{Koskinen, Malola,
  and H\"akkinen}}]{koskinen_PRB_09}
\bibinfo{author}{\bibfnamefont{P.}~\bibnamefont{Koskinen}},
  \bibinfo{author}{\bibfnamefont{S.}~\bibnamefont{Malola}}, \bibnamefont{and}
  \bibinfo{author}{\bibfnamefont{H.}~\bibnamefont{H\"akkinen}},
  \bibinfo{journal}{Phys. Rev. B} \textbf{\bibinfo{volume}{80}},
  \bibinfo{pages}{073401} (\bibinfo{year}{2009}).

\bibitem[{\citenamefont{Meyer et~al.}(2008)\citenamefont{Meyer, Kisielowski,
  Erni, Rossell, Crommie, and Zettl}}]{Meyer_NL_08}
\bibinfo{author}{\bibfnamefont{J.}~\bibnamefont{Meyer}},
  \bibinfo{author}{\bibfnamefont{C.}~\bibnamefont{Kisielowski}},
  \bibinfo{author}{\bibfnamefont{R.}~\bibnamefont{Erni}},
  \bibinfo{author}{\bibfnamefont{M.}~\bibnamefont{Rossell}},
  \bibinfo{author}{\bibfnamefont{M.}~\bibnamefont{Crommie}}, \bibnamefont{and}
  \bibinfo{author}{\bibfnamefont{A.}~\bibnamefont{Zettl}},
  \bibinfo{journal}{Nano Lett.} \textbf{\bibinfo{volume}{8}},
  \bibinfo{pages}{3582} (\bibinfo{year}{2008}).

\bibitem[{\citenamefont{Ferrari and Robertson}(2000)}]{Ferrari_PRB_00}
\bibinfo{author}{\bibfnamefont{A.}~\bibnamefont{Ferrari}} \bibnamefont{and}
  \bibinfo{author}{\bibfnamefont{J.}~\bibnamefont{Robertson}},
  \bibinfo{journal}{Phys. Rev. B} \textbf{\bibinfo{volume}{61}},
  \bibinfo{pages}{14095} (\bibinfo{year}{2000}).

\end{thebibliography}

\end{document}